\documentclass[useAMS,usenatbib]{mn2e}

\usepackage{graphicx}
\usepackage{bm}
\usepackage{natbib}

\newcommand{\be}{\begin{equation}}
\newcommand{\ee}{\end{equation}}

\newcommand{\ba}{\begin{eqnarray}}
\newcommand{\ea}{\end{eqnarray}}

\newcommand\eg{\textit{e.g.}}

\newcommand{\Bf}{{magnetic field}}
\newcommand{\Bfs}{{magnetic fields}}

\title[Magnetized cavities and UHECR acceleration]{Dynamics of  rising magnetized cavities and UHECR acceleration in clusters of galaxies}

\author[K.~N.~Gourgouliatos \& M.~Lyutikov]{Konstantinos~N.~Gourgouliatos\thanks{E-mail: kgourgou@purdue.edu} \& Maxim~Lyutikov \\
 Department of Physics, Purdue University, \\
 525 Northwestern Avenue,
West Lafayette, IN
47907-2036 } 

\begin{document}

\date{Accepted -. Received -; in original form -}
\pagerange{\pageref{firstpage}--\pageref{lastpage}} \pubyear{-}
\maketitle

\label{firstpage}

\begin{abstract}
We study the expansion of low density cavities produced by  Active Galactic Nuclei jets in clusters of galaxies. The long term stability of these cavities requires the presence of linked  magnetic fields. We find  solutions describing the  self-similar expansion of  structures containing large-scale electromagnetic fields. Unlike the  force-free spheromak-like configurations, these solutions have no surface currents and, thus, are less susceptible to resistive decay. The cavities are  internally confined by external pressure,  with zero gradient at the surface.  If the adiabatic index of the plasma within the cavity is  $\Gamma>4/3$, the expansion ultimately leads to the formation of  large-scale current sheets. The resulting dissipation of the magnetic field can only partially offset the adiabatic and radiative losses of radio emitting electrons.  

We demonstrate  that  if the formation of  large-scale current sheets is accompanied by  explosive reconnection of the  magnetic field, the resulting reconnection layer  can accelerate  cosmic rays to ultra high energies.   We speculate that the enhanced flux of  UHECRs towards Centaurus A originates at the cavities due to magnetic reconnection.
\end{abstract}


\section{Introduction}

Active galactic nuclei (AGNs) produce relativistic jets emanating from the central black hole that shock the intracluster medium (ICM), and create  over-pressurized lobes. As these lobes expand to achieve pressure equilibrium, they  form low density, high entropy cavities, also referred to as ``bubbles". Most bubbles  are not expanding supersonically and, thus, are  in pressure balance  with  their surroundings. Their rise in the cluster potential is controlled by buoyancy, and they appear as depressions in the X-ray emissivity  \citep{Fabian:2000, Churazov:2001, Birzan:2004, Diehl:2008}. The  jets that create the bubbles contain both plasma and magnetic fields.  During the active stage the location of the jet termination is seen as a radio lobe, while  cavities are likely to be the outcome of disconnected lobes during  the quiescent stage.  Jets are active for typical timescales of $\sim 10^{7} {\rm years}$, with luminosities of $10^{44}{\rm erg~s^{-1}}$. During the quiescent stage bubbles  rise in the intergalactic medium.

It is quite likely that the magnetic field plays a crucial r\^ole, however there is no direct measurement yet. Cavities survive on timescales much longer than the timescale of buoyant rise. Thus, rising bubbles in the ICM must remain coherent, and avoid being shredded by the Rayleigh-Taylor instability (RT), Richtmyer-Meshkov instability (RM) and Kelvin-Helmholtz instability (KH). On the other hand, {\it  simple hydrodynamic models ``face multiple failures''} \citep{Reynolds:2005} when addressing bubble dynamics: in fluid simulations the RT and KH instabilities quickly disrupt the bubbles on  approximately one rise time \citep[\eg][]{Kaiser:2005}. One of the principal reasons, perhaps,  for the  failure of fluid codes is that the importance of magnetic fields has been generally underestimated in the ICM, based on the notion that magnetic fields  are dynamically subdominant. In fact, {\it  even dynamically subdominant  \Bf\ can have drastic effects on bubble evolution, stability and entropy production}. First, magnetic fields in the ICM drape around expanding cavities forming a layer of nearly equipartition \Bf, which stabilizes the bubble against the Kelvin-Helmholtz and Rayleigh-Taylor instability \citep{Dursi:2007,Dursi:2008}. Secondly, if the AGN jet which blew the bubble carries large-scale magnetic fields, it will also stabilize the bubble. Stable magnetic models for cavities have been proposed by \cite{Benford:2006, Dong:2009, Gruzinov:2010, Braithwaite:2010}, who find the fields inside the bubble and suppress instabilities. Examples of magnetic fields in the exterior of the bubble that contribute to the stability of  the  cavities via the magnetic draping effect have been presented by \cite{Lyutikov:2006}. Nevertheless, it has been argued that unmagnetised bubbles can be stable provided a dense shell forms around it during the inflation stage \citep{Sternberg:2008}. 

Motivated by the above astrophysical structures we address and solve the following physical problem. Let us assume that a magnetic structure containing plasma and magnetic field in equilibrium is confined by some external pressure. This pressure is taken to be constant on the boundary of the cavity. As the bubble rises in the gravitational potential of a cluster,  the external pressure decreases. How will the internal structure of the bubble adjust to the changing external conditions? We find that the evolution of the cavity depends on the adiabatic index of the material: for $\Gamma=4/3$ the cavity expands self-similarly keeping the ratio of the kinetic plasma pressure over magnetic pressure constant, while for $\Gamma> 4/3$  the structure will be  dominated by  the magnetic field pressure and by kinetic plasma pressure for $\Gamma<4/3$. As a  result, for $\Gamma >4/3$ the expansion of the cavity  will lead to the spontaneous formation of electric current sheets, which are subject to  resistive decay of the magnetic field, i.e. by forming large-scale reconnection layers.

To study the evolution of the cavity we use the result of previous work by \cite{GBL:2010} (hereafter GBL), who have shown that it is possible to find analytical spheromak-type solutions of the structure of a magnetic field of spherical geometry, without surface currents, at the expense of including some plasma pressure. Those solutions are static and stable under MHD instabilities, while the pressure on their surface is constant allowing them to come to equilibrium with a constant pressure environment, without any deformation at all. As current sheets occur only when there is a discontinuity, those structures have no currents sheets and are not susceptible to resistive instabilities, thus magnetic flux conservation is a reasonable assumption.

We study the expansion of the cavity by applying a self-similar solution. Substantial work has been done in self-similarly expanding magnetic structures \citep{Low:1982, Prendergast:2005, Tsui:2007, GL:2008, GV:2010, Takahashi:2011}. \cite{LG:2011} and \cite{Dalakishvili:2011} have found solutions of force-free magnetic fields that expand in a self-similar manner. These structures are generalisations of spheromak and Lundquist fields, and while having exactly the same geometry, they depend on time and evolve self-similarly. In this work we apply the idea of self-similarity as described in the previous papers to the GBL solutions and we study their evolution explicitly in time. For cases where the internal dynamic timescale is shorter than the time the external parameters need to change substantially we assume quasi-static evolution: a series of equilibria solutions which satisfy the changing boundary conditions.

In addition to bubble dynamics another astrophysical process that is related to the action of AGNs is that of Ultra High Energy Cosmic Ray (UHECR) acceleration. One of the most important results of the newly commissioned  Pierre Auger Observatory is  the discovery of a correlation  between the arrival direction of UHECRs and AGNs listed in the Veron-Cetty \& Veron  catalog \citep{Abraham, Alvarez00}. In particular a number of events seem to be associated with Centaurus A, the nearst AGN. Centaurus A is  FR I radio galaxy, with a fairly low bolometric radio luminosity of $ 4 \times 10^{41}$ erg s$^{-1}$ \citep{AlexanderLeahy87}. There is a mildly relativistic jet, detected  in both radio and at high energies, emitting $\sim 5 \times 10^{42}{\rm erg~ s}^{-1}$ in keV -- MeV range \citep{Kraft02,Hardcastle03}. The orientation $\theta$ of the jet of Centaurus A with respect to our line of sight is $35 ^\circ < \theta  < 72 ^\circ$ \citep[\eg][]{Skibo:1994}. The jet is launched by a central black hole of mass $M_{BH}  \sim 4 -10 \times 10^7 M_\odot$. The galaxy is surrounded by a set of   AGN-blown cavities, inner ($\sim 1-2 $ kpc in size), intermediate ($\sim 10-20 $ kpc in size) and largescale of several hundred kiloprsecs, \cite[see, \eg][for review]{Israel:1998}. 

A crucial question is how these cosmic rays are accelerated. \cite{Giannios:2010} has suggested that reconnection can accelerate UHECRs up to the essential energies in AGN jets; \cite{Benford:2008} have suggested that this process takes place in fossil AGN jets through slow reconnection. Our suggestion in this paper is that the change of the structure of the cavity leads to the formation of current sheets, where explosive reconnection is triggered and acceleration of UHECRs actually takes place therein. An important issue with cosmic ray acceleration through reconnection is to have slow enough magnetic field dissipation at early stages so that a strong field builds, and then a physical mechanism to trigger reconnection and accelerate them. The description of the cavity evolution we propose provides these properties. They start without currents sheets, thus the magnetic field gradient is low enough to have a slow dissipation rate, but as they expand their plasma $\beta$ drops low enough to create a magnetically dominated region and a current sheet.  

The structure of the paper is the following: We start from dimensional arguments about the magnetic and plasma pressure evolution during the expansion of a magnetized cavity; then we solve explicitly the partial differential equations describing the problem: we find an exact dynamical solution and a series of quasi-static solutions which correspond to a wider range of parameters. Finally we apply our results for the bubble evolution in the acceleration of UHECRs.  

\section{Dimensional arguments}

Let us start with a sphere of radius $R$ containing plasma of total mass $M$ and and some magnetic flux $\Psi$, the plasma pressure $p$ is related to the density $\rho$ through an adiabatic index $\Gamma$. The density of the cavity shall be $\rho \propto M R^{-3}$, the magnetic field $B\propto \Psi R^{-2}$, the magnetic pressure $p_{mag} \propto \Psi^{2} R^{-4}$ and the plasma pressure $p \propto \rho^{\Gamma} \propto (M R^{-3})^{\Gamma}$. Let us consider some change by $\Delta R$ of the radius of the cavity, while the total mass and flux are conserved, and estimate the change caused to the plasma pressure, to the magnetic pressure and the plasma $\beta=p/p_{mag}$. 
\begin{eqnarray}
\frac{\Delta p}{p}=-3\Gamma \frac{\Delta R}{R}\,, \nonumber \\
\frac{\Delta p_{mag}}{p_{mag}}=-4\frac{\Delta R}{R}\,, \nonumber \\
\frac{\Delta \beta}{\beta}=(4-3\Gamma)\frac{\Delta R}{R}\,.
\end{eqnarray}
From the above result we expect that if $\Gamma <4/3$ the gas pressure will dominate over the magnetic pressure when the sphere expands, whereas if $\Gamma>4/3$ the magnetic pressure will dominate over the gas pressure after expansion, while for $\Gamma=4/3$ the ratio of pressures shall not be affected.  

For $\Gamma>4/3$, decreasing the plasma $\beta$ implies that at some point during expansion, the minimal pressure inside the sphere becomes zero. Further expansion leads to the appearance of current sheets.

\section{Magnetic field structure of the cavity}

\subsection{The static cavity}

Stability arguments require that the internal structure of \Bfs\ should be a combination of toroidal and poloidal components, which, in order to be in equilibrium with a constant external pressure, must vanish on the boundary. These conditions lead to an overdetermined mathematical problem: We seek solutions of an elliptical partial differential equation, the Grad-Shafranov equation \citep{Shafranov:1966}, with a given value of the function we are solving for and its derivative on the boundary. Thus the solution must satisfy both Newman and Dirichlet boundary conditions simultaneously. However, two more functions appear in the elliptical partial differential equation: a function related to the poloidal current and a function related to the plasma pressure. The only constraint they must satisfy is to be functions of the flux. As the pressure and the poloidal electric current are not  determined {\it a priori}, they can be chosen so that the solution satisfies the over-constraining boundary condition. This problem has been solved in GBL, and we use this result as the initial state for the rising magnetic cavity. 

\begin{figure}
	\centering
		\includegraphics[width=0.45\textwidth]{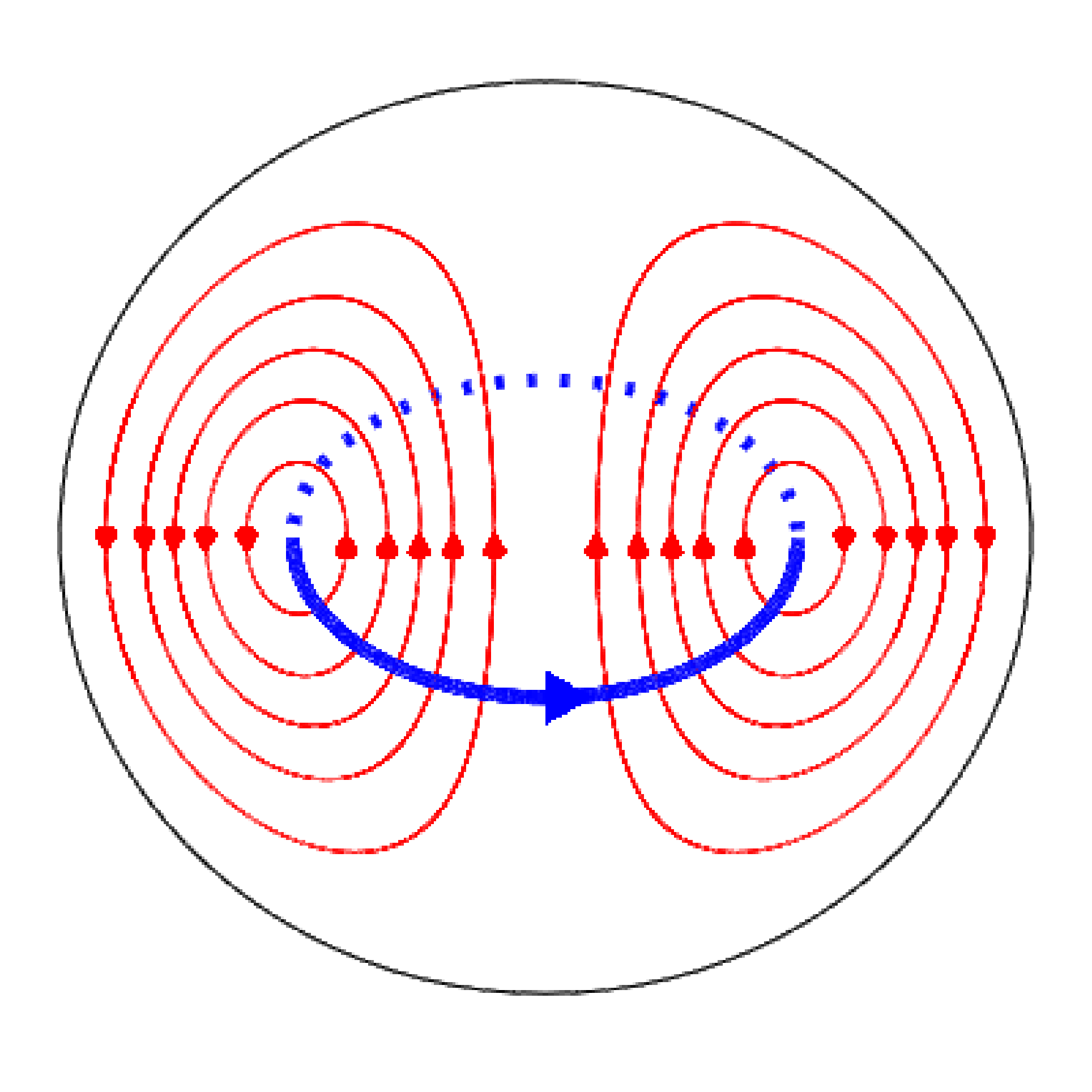}
		\caption{Plot of the the poloidal field lines (thin red) and the central toroidal field line (thick blue).}
		\label{FIELDLINES}
\end{figure}

In general the magnetic field can be expressed in terms of poloidal and toroidal components:
\begin{eqnarray}
\bm{B}=\nabla \Psi \times \nabla \phi + 2I(\Psi)\nabla \phi\,.
\end{eqnarray}
The poloidal flux and current passing through a spherical cap of radius $r$ and opening angle $\theta$ are $2\pi \Psi(r, \theta)$ and $cI(\Psi)$ respectively. A solution satisfying the Grad-Shafranov equation $\bm{J} \times \bm{B}=\nabla p$ is 
\begin{eqnarray}
\Psi=\sin^{2}\theta\Big\{c_{1}\Big[\alpha\cos(\alpha r)-\frac{\sin(\alpha r)}{r}\Big]-\frac{F_{0}}{\alpha^{2}}r^{2}\Big\}\,, \\
I=\frac{\alpha}{2} \Psi\,, \\
p=\frac{F_{0}\Psi}{4 \pi}+p_{0}\,,
\end{eqnarray} 
\begin{eqnarray}
B_{r}=\frac{2\cos\theta}{r^{2}}\Big\{c_{1}\Big[\alpha\cos(\alpha r)-\frac{\sin(\alpha r)}{r}\Big]-\frac{F_{0}}{\alpha^{2}}r^{2}\Big\}\,, \nonumber \\ 
B_{\theta}=\frac{\sin\theta}{r}\Big\{c_{1}\Big[\frac{a\cos(\alpha r)}{r}-\sin(\alpha r)\Big(\frac{1}{r^{2}}-\alpha^2\Big)\Big]+\frac{2F_{0}r}{\alpha^{2}}\Big\}\,, \nonumber \\
B_{\phi}=\frac{\alpha \sin\theta}{r}\Big\{c_{1}\Big[\alpha\cos(\alpha r)-\frac{\sin(\alpha r)}{r}\Big]-\frac{F_{0}}{\alpha^{2}}r^{2}\Big\}\,.
\end{eqnarray}
This solution has been studied in detail in GBL, and corresponds to a linear relation between the physical quantities: poloidal magnetic flux, poloidal electric current and plasma pressure. Given an appropriate choice of $\alpha$ and $F_{0}$ the field is confined inside a spherical cavity (Figure~\ref{FIELDLINES}) and the surface currents can be eliminated, while the whole structure is stable . In the static form of the Grad-Shafranov equation the pressure appears through its derivative, while $p_{0}$ is a constant which is used so that there is no negative energy density inside the cavity, thus it has a minimum permitted value, but no upper limit. The cavity is a dip in plasma pressure and density compared to the external medium, whereas the magnetic field inside the cavity is higher. The choice of $p_{0}$ affects $\beta$, the ratio of the plasma pressure over the magnetic pressure in the cavity. The value of $\beta$ is not constant within the cavity, we can evaluate an average value 

\begin{eqnarray}
\bar{\beta}=\frac{\int p_{mag} dV}{\int p dV}\,. 
\end{eqnarray}
The value for $\bar{\beta}$ that corresponds to the minimum permitted value of $p_{0}$ is $0.8$, however this is mainly due to the contribution of the area near the boundaries of the cavity, if we stop the integration at $0.9$ of the radius then the value of $\bar{\beta}$ drops to $0.5$, while if we stop at $0.7$ the value of $\bar{\beta}$ is as low as $0.2$.   

The values for $\alpha$ and $F_{0}$ are given subject to the conditions that $\Psi(r_{c})=0$ and $\frac{d\Psi}{dr}|_{r_{c}}=0$, where $r_{c}$ is the radius of the sphere, these conditions lead to zero fields on the surface and a smooth transition from the cavity to the external medium. For instance, setting $r_{c}=1$, and normalising the maximum value of $\Psi$ to unity, we find that $\alpha=5.76$, $c_{1}=-0.134$ and $F_{0}=-24.46$.

It is important to point out that although this solution is topologically similar to the spheromak field which is a force-free structure inside a sphere, there is a crucial difference regarding the surface pressure. The pressure on the surface of a spheromak is anisotropic. It is zero on the pole and reaches a maximum on the equator. Thus a spheromak cannot be in equilibrium with a constant pressure environment. On the contrary, our solution can be in a stable equilibrium with such an environment.

\subsection{Self-similar expansion}

In this section we allow $\alpha=\alpha(t)$ and we introduce time-dependence through self-similar expansion of concentric shells with velocity $\bm{v}=-r\frac{\dot{\alpha}}{\alpha}\hat{\bm{r}}$. The evolution of the structure is given by the momentum equation, while it must satisfy Maxwell equations, the continuity equation and the entropy equation. Given that we are working in a non-relativistic context we shall use the Maxwell equations keeping only the appropriate terms. 
\begin{eqnarray}
\nabla \cdot \bm{B}=0\,, \\
\nabla \times \bm{E}=-\frac{1}{c}\frac{\partial \bm{B}}{\partial t}\,, \\
\nabla \times \bm{B}=\frac{4\pi}{c}\bm{j}\,,
\end{eqnarray}
We shall not consider the dynamical effects of the electric field, thus we do not take into account Gauss' law for the electric field and in the Amp\`ere-Maxwell equation we shall not take into account the displacement current. The electric field through the ideal MHD approximation is
\begin{eqnarray}
\bm{E}=-\frac{\bm{v}}{c}\times \bm{B}\,.
\end{eqnarray}
The continuity equation is 
\begin{eqnarray}
\frac{\partial \rho}{\partial t} +\nabla \cdot (\rho \bm{v} )=0\,.
\end{eqnarray}
The momentum equation is
\begin{eqnarray}
\rho\Big(\frac{\partial}{\partial t} +\bm{v} \cdot \nabla \Big)\bm{v}=-\nabla p +\frac{\bm{j} \times \bm{B}}{c}\,.
\label{MOM}
\end{eqnarray}
Finally the entropy equation is
\begin{eqnarray}
\Big(\frac{\partial}{\partial t} +\bm{v} \cdot \nabla\Big)\Big(\frac{p}{\rho^{\Gamma}}\Big)=0\,.
\label{ENTROPY}
\end{eqnarray}
Flux and helicity conservation under self-similar expansion lead to the following form for $\Psi$
\begin{eqnarray}
\Psi=\frac{\sin^{2}\theta }{\alpha_{0}^{2}}\Big\{c_{1}\Big[\frac{\sin(\alpha(t) r)}{\alpha(t)r}-\cos(\alpha(t)r)\Big]-F_{0}(\alpha(t)r)^{2}\Big\}\,.
\label{SOLUTION}
\end{eqnarray}
Note that $\Psi=\Psi(\alpha(t)r, \theta)$. The boundary conditions, similarly to the static case, are that the magnetic flux and its derivative are both zero at the boundary of the cavity $r_{c}$, while the boundary of the cavity now depends on time $r_{c}=\frac{\alpha_{0} r_{0}}{\alpha(t)}$. The field is expressed in terms of a poloidal and a toroidal component. 
\begin{eqnarray}
\bm{B}=\nabla \Psi \times \nabla \phi +\alpha(t)\Psi\nabla \phi\,,
\end{eqnarray}
Constraining our interest to uniform expansion without acceleration we take from the left-hand-side of the momentum equation~(\ref{MOM}) that $\alpha(t) =(v_{0}t+r_{0})^{-1}$, where $v_{0}$ is the expansion velocity of the boundary of the cavity. The plasma pressure becomes
\begin{eqnarray}
p=p_{0}+\frac{F_{0}\alpha(t)^{4} \Psi}{4 \pi \alpha_{0}^{2}}\,.
\label{PRESSURE}
\end{eqnarray}
From the baryonic mass conservation equation we find that the density is 
\begin{eqnarray}
\rho(r, \theta, t)=\tilde{\rho}(\alpha(t) r, \theta)(\alpha(t)r_{0})^{3}\,,
\label{DENSITY}
\end{eqnarray}
where $\tilde{\rho}$ is a function of the self similar variable and $\theta$. From the entropy equation (\ref{ENTROPY}) it is $p=K\rho^{\Gamma}$ and combining this with equation~(\ref{DENSITY}) we find for $\Gamma=4/3$ that $p_{0}$ and $\alpha$ are related through $p_{0}(t)\propto \alpha(t)^{4}$.

\subsection{Quasi-static evolution}
\label{QUASISTATIC}

In the previous section we have found an exact solution for the evolution of the cavity. However, this solution constrains the value of $\Gamma$ and the time dependence of $p_{0}$. In this section we investigate the evolution of the cavity when the external parameters change slowly compared to the dynamic timescale of the cavity itself. Under this assumption we can describe the evolution as a series of static solutions which are subject to the changing boundary conditions \citep{Lynden-Bell:2003}, without constraining $\Gamma$, or requiring any time dependence for $p_{0}$. Such solutions need to satisfy the right hand side of equation~(\ref{MOM}), and as the evolution is slow inertia terms are negligible; practically we find Grad-Shafranov equilibria. 

From equations~(\ref{PRESSURE}), (\ref{DENSITY}) and the adiabatic relation ($p \propto \rho^{\Gamma}$) we find that 
\begin{eqnarray}
p_{0}+\frac{F_{0}\alpha^{4} \Psi}{4 \pi \alpha_{0}}=K (\tilde{\rho} r_{0}^{3} \alpha^{3})^{\Gamma}\,.
\end{eqnarray}
Let us express $p_{0}=\tilde{p}_{0}\alpha^{n}$, where $\tilde{p}_{0}$ is constant. Taking the logarithm of the above equation we find that there is some relation between $n$ and $\Gamma$, which changes as the cavity expands and cannot be  expressed analytically, except for the special case of $\Gamma=4/3$, where $n=4$ and the radius of the cavity is $r_{c}\propto p_{0}^{-1/4}$ which leads us back to the solution of section 3.2.

To find out the approximate form of $n$ when $\Gamma\neq 4/3$, we start from a structure of a given $\bar{\beta}$ and then we decrease the pressure, demanding that the system remains in equilibrium while the mass, the flux and the helicity are conserved. Thus the solution shall be of the form of equation~(\ref{SOLUTION}). Doing so for various values of $\Gamma$ we find how the radius increases as $p_{0}$ decreases. In accordance with the dimensional discussion of section 2, there are three distinct regimes depending on whether $\Gamma$ is greater, smaller or equal to $4/3$. 

\begin{figure}
	\centering
		\includegraphics[width=0.45\textwidth]{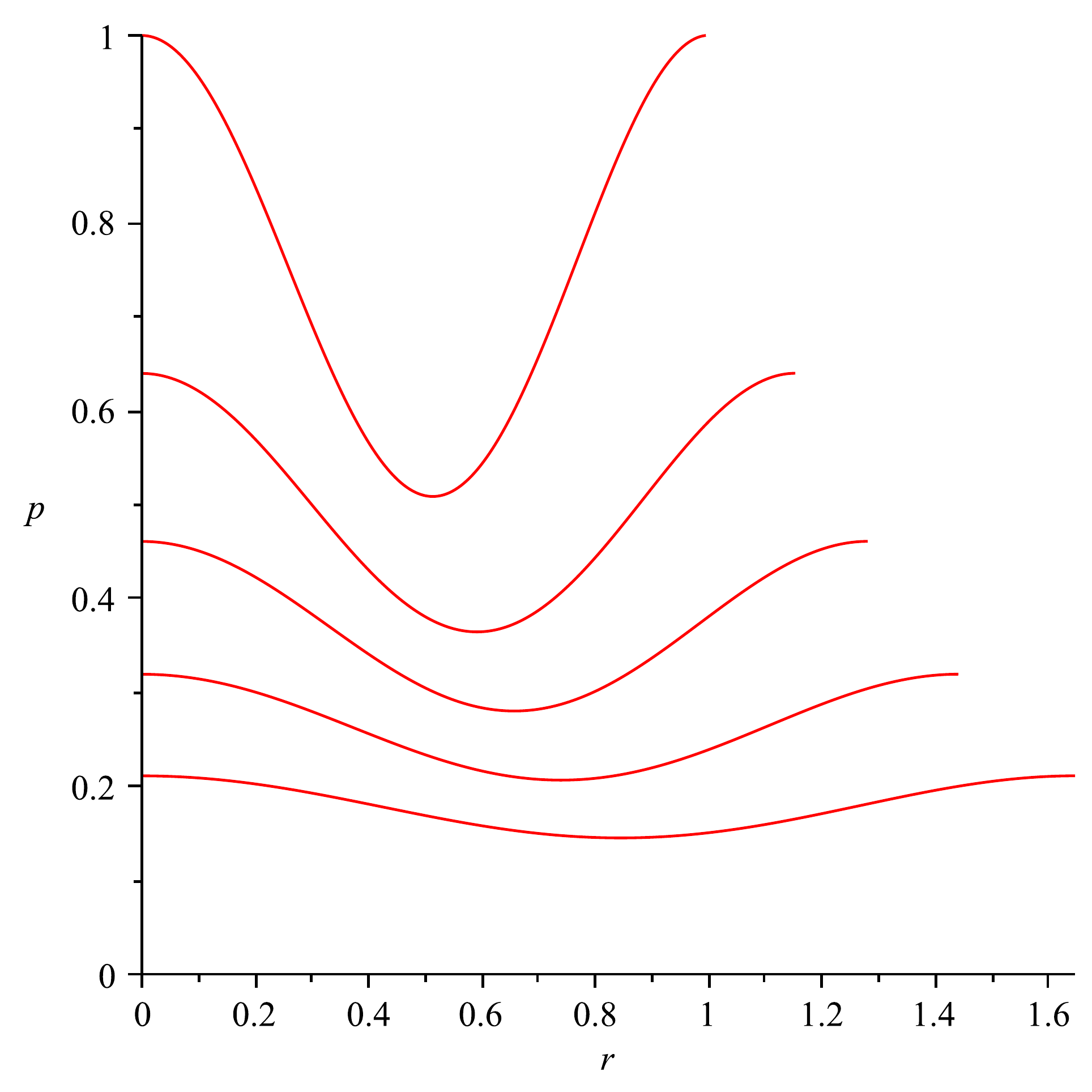}
		\caption{The pressure for the expanding cavity with $\Gamma=1$, presented as a paradigm of pressure with smaller adiabatic index than $4/3$. The cavity expands while $p_{0}$ drops. As it expands the gas pressure dominates over the magnetic one and finally the magnetic cavity gets diluted.}
		\label{F_1}
\end{figure}

\begin{figure}
	\centering
		\includegraphics[width=0.45\textwidth]{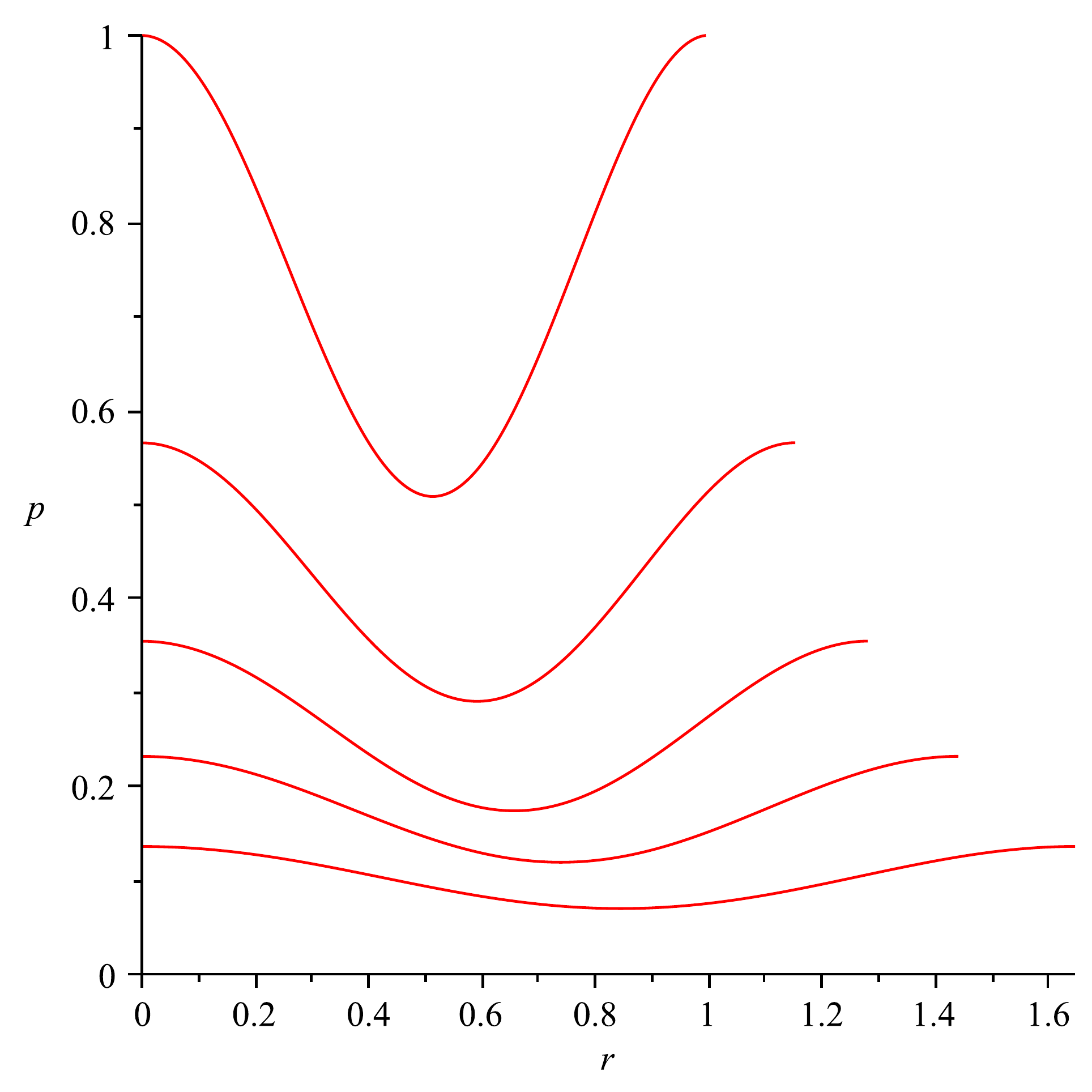}
		\caption{The pressure for the expanding cavity with $\Gamma=4/3$. The cavity expands while $p_{0}$ drops, but it does not become diluted nor does it reach zero density at the centre.}
		\label{F4_3}
\end{figure}

\begin{figure}
	\centering
		\includegraphics[width=0.45\textwidth]{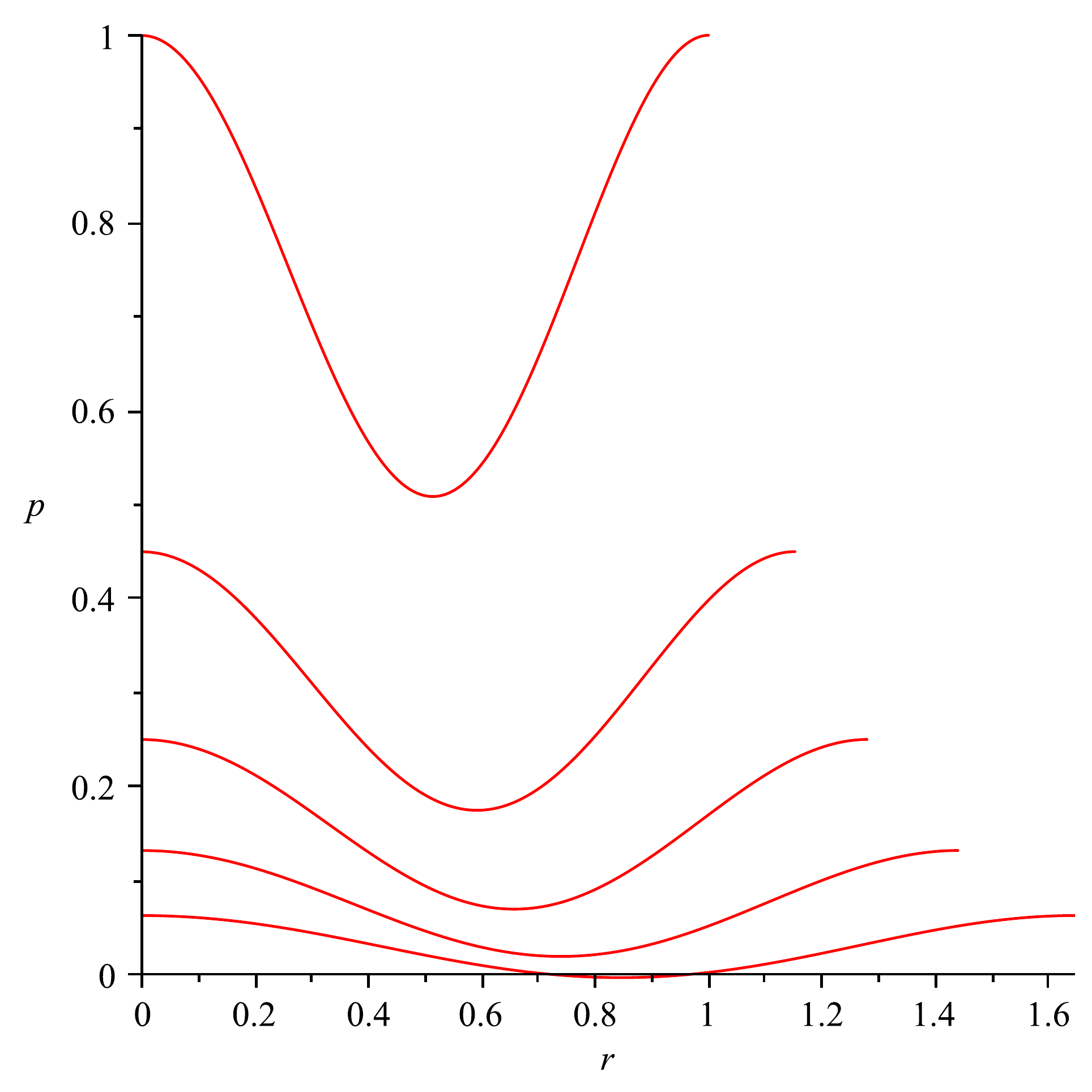}
		\caption{The pressure for an expanding cavity with $\Gamma=2$, presented as a paradigm for plasma with $\Gamma>4/3$. The cavity expands while $p_{0}$ drops, after some amount of expansion the density at the centre becomes zero, while the field there is no longer zero. At this stage the field in the centre of the cavity tries to reach a force-free equilibrium while the rest of the cavity has a pressure-magnetic field equilibrium. This state cannot be described by the solution presented in this paper.}
		\label{F_2}
\end{figure}

\begin{figure}
	\centering
		\includegraphics[width=0.45\textwidth]{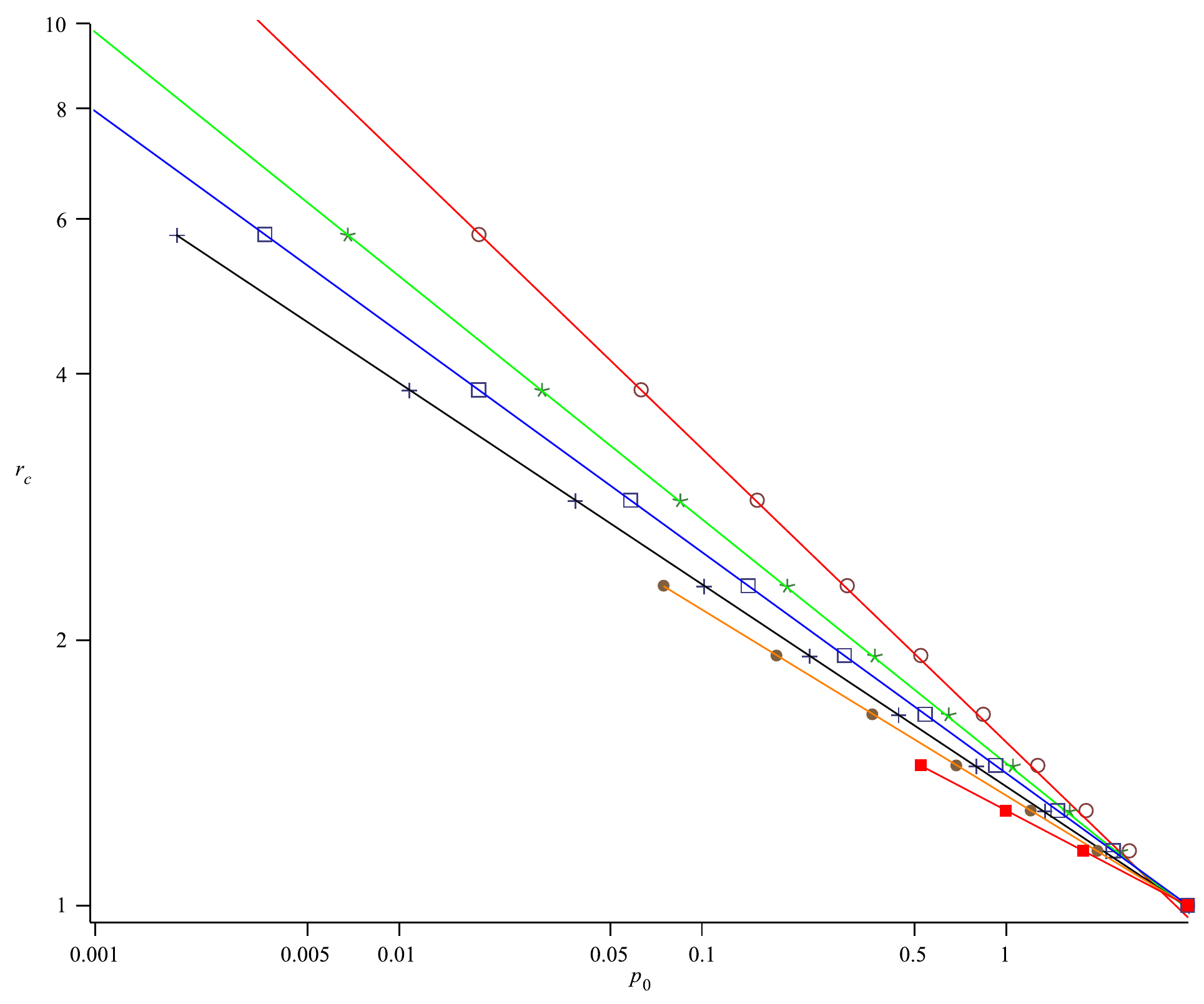}
		\caption{The dependence of the radius of the cavity $r_{c}$ on $p_{0}$, for various values of $\Gamma$, empty circles are for $\Gamma=1$, asterisks $\Gamma=6/5$, empty squares $\Gamma=4/3$, cross $\Gamma=3/2$, solid circle $\Gamma=5/3$ and solid square $\Gamma=2$. The dependence of $r_{c}$ on $p_{0}$ can be approximated by a power-law. For $\Gamma>4/3$ the lines cannot be extrapolated to arbitrarily low pressure, after the point that corresponds to maximum expansion the minimum pressure in the cavity goes to zero and further expansion leads to the formation of surface currents. }
		\label{p0_rb}
\end{figure}

\begin{figure}
	\centering
		\includegraphics[width=0.45\textwidth]{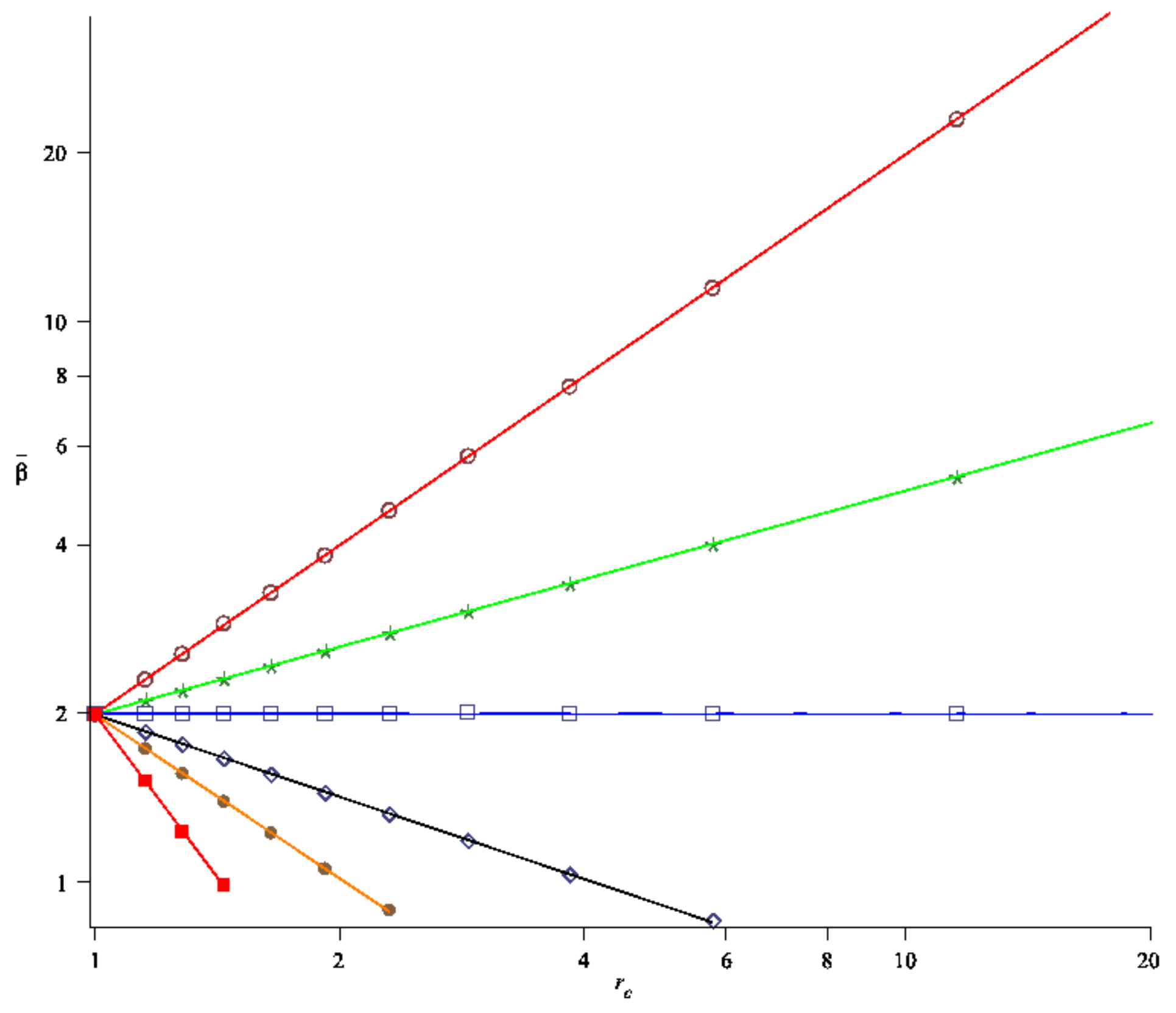}
		\caption{The plasma $\bar{\beta}$ as a function of $r_{c}$, for various values of $\Gamma$, empty circles are for $\Gamma=1$, asterisks $\Gamma=6/5$, empty squares $\Gamma=4/3$, cross $\Gamma=3/2$, solid circle $\Gamma=5/3$ and solid square $\Gamma=2$. The dependence can be approximated by a power law. We find that for $\Gamma>4/3$ and if $\bar{\beta}\sim 0.8$ the minimum plasma pressure in the cavity becomes zero and further expansion leads to the formation of current sheets. }
		\label{rb_beta}
\end{figure}

If $\Gamma=4/3$, $\bar{\beta}$ does not change with expansion. This is not the case when $\Gamma>4/3$, since the expansion causes a faster decrease of the plasma pressure over the magnetic, leading to smaller $\bar{\beta}$. At some critical point, the gas pressure at the minimum becomes zero so that the evolution under the self-similar scheme is impossible. At this stage we expect that reconnection becomes important. If $\Gamma<4/3$, as the cavity expands its magnetic pressure becomes weaker compared to the gas pressure while $\bar{\beta}$ increases, and finally the magnetic field becomes diluted in the background, see Figures~(\ref{F_1})-(\ref{F_2}). The inverse process takes place if we study the effect of compression instead. From this discussion we can determine the dependence of $r_{c}$ on $p_{0}$. The general result is plotted in Figure~(\ref{p0_rb}), we find $r_{c} \propto p_{0}^{n}$, where $n\approx -(3\Gamma)^{-1}+\epsilon_{n}$, $\Gamma$ appears because of the plasma content of the cavity, while $\epsilon_{n}$ is related to the contribution of the magnetic pressure in the system and its value is such that $n$ will tend towards $-1/4$, with greater contribution when the magnetic pressure is important in the cavity. In a similar way we evaluate the evolution of $\bar{\beta}$ under expansion for various values of $\Gamma$. From Figure~(\ref{rb_beta}) we find that $\bar{\beta}$ obeys a power law with $\bar{\beta}\propto r_{c}^{m}$, where $m=4-3\Gamma+ \epsilon_{m}$; $\epsilon_{m}$ is a parameter so that $m$ will tend towards $0$ for strongly magnetized structures. In general $\epsilon_{n}$ and $\epsilon_{m}$ will be small compared to $n$ and $m$. Finally we conclude that $\bar{\beta} \propto r_{c}^{-4/(3\Gamma)+1+\epsilon_{p}}$ where $\epsilon_{p}=(4-3\Gamma)\epsilon_{n}-\frac{1}{3\Gamma}\epsilon_{m}$. 

\section{Acceleration of UHECR through explosive reconnection}

In this section  we shall do some order of magnitude calculations to show that it is possible to reach the essential potentials for the UHECR acceleration. For the numerical estimates below we assume that the intrinsic  Centaurus A jet power is  $L_j=10^ {44}L_{44}$ ergs s$^{-1}$, while during flares it can reach $10^ {45}$ ergs s$^{-1}$. We also assume that the central black hole mass  is  $M_{BH}  \sim 10^8 M_\odot$ and  jet activity lasted for $\sim 10 $ Myr, close to the estimate of \cite{Hardcastle09} of 30 Myr. We shall show that the weak reconnection throughout the evolution of the cavity is not sufficient to reach the available potentials. However, as we have shown in the previous section, a cavity with initial $\bar{\beta}>1$ and $\Gamma>4/3$, after some expansion will have parts with zero plasma pressure and cannot evolve further in this regime. At this stage something must happen, e.g. formation of current sheets, which become resistive and form reconnection layers.

\subsection{Unmagnetised cavity dynamics}

The jet ejects material into the intergalactic medium with particle density $10^{-5}{\rm cm^{-3}}$. The cavity formed by ejected material is confined within a radius $r_{c}$, further out is the forward shock with radius $r_{FS}$ formed from the compressed IGM material. From the Sedov-Taylor solution the forward shock radius is
\begin{eqnarray}
r_{FS}\sim \frac{L^{1/5}t^{3/5}}{\rho_{IGM}^{1/5}}\sim 60{\rm kpc}~L^{1/5}_{44}t^{3/5}_{7}n^{-1/5}_{-5} \,.
\end{eqnarray}
As the cavity is in equilibrium with the shocked IGM, the pressure of the cavity $\rho_{c}^{\Gamma}$ and that of the shocked IGM $\rho_{IGM}(r_{FS}/t)^{2}$ are equal. Assuming that most of the energy comes in the form of mildly relativistic ejecta $L\sim \dot{M}c^{2}$, we find for $\Gamma=4/3$ that 
\begin{eqnarray}
r_{c}\sim \Big(\frac{L^{7}t^{16}}{c^{5}\rho_{IGM}}\Big)^{1/30}\sim 30{\rm kpc} ~ L^{7/30}_{44}t^{8/15}_{7}n^{-7/30}_{-5}\,,
\end{eqnarray}
and the ratio of the two radii is
\begin{eqnarray}
\frac{r_{c}}{r_{FS}}=0.5 ~ L^{1/30}_{44}t^{-1/15}_{7}n^{-1/30}_{-5}\,.
\end{eqnarray}
which is  a very weak function of the parameters, thus the cavity shall be about half the size of the forward shock, unless there is an extremely drastic change of the physical parameters of the problem. 

\subsection{Evolution of a weakly resistive magnetised cavity}
\label{SLOWREC}

Let us assume that a fraction $\sigma_{j} \leq 1$ of the jet luminosity comes out in the form of a large scale magnetic field. Then, the jet injects a magnetic flux $\Psi$ at a rate $(\sigma_{j} L c)^{1/2}$. At the same time magnetic flux is destroyed by reconnection at a rate $\frac{\Psi}{t_{r}}$, where $t_{r}$ is the reconnection timescale. Thus the evolution of the flux in the cavity obeys
\begin{eqnarray}
\dot{\Psi}=(\sigma_{j} L c)^{1/2}-\frac{\Psi}{t_{r}}\,.
\end{eqnarray}
We expect that the magnetic field within the cavity will relax to a minimum energy state on an Alfv\'en time  scale, which as we show below can be very short. Such a minimum energy state will include a toroidal-poloidal configuration of magnetic field as a purely toroidal or poloidal field is unstable and the co-existence of plasma will make the solution described in the previous sections appropriate. As the AGN-blown cavity expands, it does work on the shocked IGM, so its energy is not conserved; on the other hand, in the absence of resistivity the magnetic flux is conserved. Neglecting the resistive decay of magnetic flux, the magnetic field in  the cavity is
\begin{eqnarray}
B \sim { \dot{\Psi} t \over r_{c}^2} \sim \sqrt{ \sigma_{j}}     \left( { L   c^{25} \rho_{IGM} ^{14}\over t^2}\right)^{1/30} = 60 \mu {\rm G}   \sqrt{ \sigma_{j}}    n_{-5}^{7/15}\,.
\end{eqnarray}
The density  in  the cavity is very low
\begin{eqnarray}
\rho_{c} =  { (L/c^2) t \over r_{c}^3}= \left( { L ^3  \rho_{IGM} ^{7}\over  c^{15} t^6}\right)^{1/10} = 3 \times 10^{-8} {\rm cm}^{-3}n_{-5}^{7/10}\,.
\end{eqnarray}
By equating the shocked IGM pressure to $\rho_{c} v_T^2$, we can find the typical thermal velocity of the cavity particles
\begin{eqnarray}
{v_T\over c} =  \left( L \over c^5  \rho_{IGM} t ^2 \right)^{1/20} =0.4\,. 
\end{eqnarray}  
If the jet is ion-dominated, the corresponding ion temperature is $ T_i \sim v_T^2 m_p \sim 100 $ MeV, while for pair-dominated jet, the electron temperature is $45 $ keV. 

When jet ions pass through  a reverse shock they are heated to $T_i \sim m_p c^2\sim 10^9$ eV. If electrons are coupled to ions effectively, they are heated to the same temperature. This justifies the use of adiabatic index $\Gamma=4/3$. As the cavity expands, both species cool adiabatically, the plasma pressure decreases even faster and the particles become non-relativistic, leading to $\Gamma>4/3$, so that the magnetic pressure eventually dominates the cavity. In addition, both species can be accelerated non-thermally.

The Alfv\'en velocity is
\begin{eqnarray}
{v_A  \over c}= { B \over  c \sqrt{4 \pi \rho_{c} }}= \sqrt{ \sigma_{j}}      \left( {   \rho_{IGM} ^{7} c^{35}  t^{14} \over L^7}\right)^{1/60}  = 3  \sqrt{ \sigma_{j}} \,.  
\label{vA}
\end{eqnarray}
Let $\eta_{A}$ be the reconnection efficiency, a charge ${\rm e}$ shall be accelerated to energy
\begin{eqnarray}
{\rm e} \Phi = \eta_A  {v_A  \over c} B r_{c} = \eta_A \sigma_{j} c^{5/4} L^{3/20} t^{7/10} \rho_{IGM} ^{7/20} = \nonumber \\
 2 \times 10^{22}{\rm eV} ~ \eta_A  \sigma_{j}  n_{-5}^{7/20} \,.
\end{eqnarray}
We can neglect resistive decay provided that the reconnection timescale,
\begin{eqnarray}
t_r  \sim r_{c} / ( \eta_A v_A)  = {1 \over \eta_A  \sqrt{ \sigma_{j}}  }  \left( {  L^7 t^6 \over c^{35}  \rho_{IGM} ^{7}}\right)^{1/20}\,,
\end{eqnarray}
 is longer than the AGN activity time, $t_r > t$. Thus,
 \begin{eqnarray}
 t\sim t_r> { \sqrt{L} \over ( \eta_A  \sqrt{ \sigma_{j}}   )^{10/7} c^{5/2} \sqrt{ \rho_{IGM}} }\,.
 \label{ttr}
 \end{eqnarray}
 We should also make sure that the Alfv\'en velocity, which increases with time, Eq. (\ref{vA}) does not exceed the speed of light
\begin{eqnarray}
t \leq { \sqrt{ L } \over c^{5/2} \sigma_{j} ^{30/14} \sqrt{  \rho_{IGM} }}\,.
\label{vva}
\end{eqnarray}
The most effective acceleration will occur right before the resistivity becomes important. Under the constraint that the Alfv\'en velocity at this point is smaller than the speed of light, Eqns (\ref{ttr}-\ref{vva}) give $ \sigma_{j} = \eta_A$.  The resistive time then becomes
\begin{eqnarray}
t_r = { \sqrt{L} \over  \eta_A ^{15/7} c^{5/2} \sqrt{ \rho_{IGM}} } \,.
\label{trr1}
\end{eqnarray}
and
\begin{eqnarray}
\eta_A =  \left( {  L^7  \over c^{35} t_{AGN}^{14} \rho_{IGM} ^{7}}\right)^{1/30} = 10^{-2} L_{44}^{7/30}  t_{7}^{-7/15} n_{-5}^{-7/30}\,.
\end{eqnarray}
This value leads to a jet where about $10^{-2}$ of its energy budget comes in as magnetic field, thus it is consistent with the total luminosity of the system.  
Using this estimate of $\eta_A$ and the condition $\sigma_{j} = \eta_A$, the available potential shall accelerate a charge ${\rm e}$ to energy 
\begin{eqnarray}
{\rm e} \Phi =   \left( {  L^{37} \over c^{65}  \rho_{IGM} ^{7} t_{AGN}^{14} }\right)^{1/60} =   2 \times 10^{18} {\rm eV}\,. 
\label{phi}
\end{eqnarray}
The main reason the potential (\ref{phi}) is fairly small is that effective reconnection requires high reconection speeds ($\Phi \propto v_{rec}$), but if reconnection is too efficient the magnetic  field is destroyed too early ($\Phi \propto B $). Still, it might be a viable mechanism. First, our order-of-magnitude estimates can easily miss a factor of a few; in addition,  accelerated particles can be of CNO (Centaurus A is located sufficiently close) or Fe type. In that case, the acceleration shall take place in the area where the field reverses through slow reconnection as it has been proposed by \cite{Benford:2008}.

\subsection{Explosive reconnection}

In section \ref{SLOWREC} we have shown that steady reconnection cannot lead to sufficient particle acceleration for UHECRs  since the magnetic field constantly decays, reducing the available potential. However, reconnection in astrophysics  is rarely a steady state process. For instance, solar flares can develop on timescales of minutes, while being driven, one might argue, on the solar  dynamo of 22 years. 
 
What we have shown in section \ref{QUASISTATIC} is that quasi-static evolution with the appropriate $\Gamma$ leads to the formation of a region with very high magnetization which shall give a high value for $\eta_{A}\sim 1$ as a strongly magnetized area forms inside the cavity. This drastic change leads now to a new estimation for the available potential. 
\begin{eqnarray}
{\rm e} \Phi = \eta_A {v_A \over c} B r_{c}  = \eta_A  c^{5/4} L^{3/20} t^{7/10} \rho_{IGM}^{7/20}=  \nonumber \\ 
=2  \times 10^{20} L_{44}^{3/20}  t_{7}^{7/10} n_{-5}^{-7/20}~ {\rm eV} \,.
\label{POTENTIAL}
\end{eqnarray}
As the bubble with $\Gamma > 4/3$ expands, at some moment a current sheet is formed. We suggest that this serves as a reconnection trigger. The ensuing reconnection sets large scale flows towards the dissipation region. Equation~(\ref{POTENTIAL}) then gives the total electric potential  of the reconnection layer. This potential is sufficiently high to account for the acceleration of UHECRs.

\cite{Alvarez00} has found that the distribution of UHECRs is not isotropic with a confidence level of 99\%. These observations suggest that there is some correlation with AGN sources, especially with Centaurus A, which is a radio galaxy that lies within the GZK horizon. Centaurus A hosts fossil radio bubbles that are the final stages of evolution of the bubbles blown by AGNs \citep{Israel:1998}. In addition, since it is very close, the CRs may comprise of CNO-type nuclei; this decreases the requirements on the available potential. 

One may expect two types of acceleration processes in reconnection regions. First, there is acceleration by DC electric field within the dissipative region itself. This type of acceleration is unlikely to produce UHECRs since  electrons would be accelerated to sufficiently high energies and  produce pairs, which would annul the electric field. Secondly, a Fermi-I type acceleration can occur in the ideal region surrounding the reconnection layer \citep{Lazarian:1999,Giannios:2010}. 

\section{Radio luminosity}

The presence of magnetic fields in cavities leads to synchrotron radio emission; indeed some X-ray cavities have enhanced radio emission \citep{McNamara:2000, Fabian:2000} which is stronger when the cavity is closer to the AGN. During the explosive reconnection stage, it is possible to have additional ongoing acceleration within the cavities, offsetting the effects of adiabatic and radiative cooling. However, as we shall show below, this effect averaged over a longer time does not change significantly the radio profile of the cavities.

The scenario we consider is the following: When the minimum plasma pressure inside the cavity reaches zero the ideal expansion cannot proceed any more, current sheets form which are susceptible to resistive decay. Resistive decay proceeds slowly at the beginning, with $\eta_{A}\ll 1$, until the formation of current sheets with $\eta_{A} \sim 1$. The timescale for reconnection at the current sheets can be evaluated from equation (\ref{trr1}); it is $t_{r}\sim 10^{5}$. The flux available for reconnection is provided with a rate directly related to the expansion of the cavity which is longer than the reconnection timescale. Indeed, in $10^{5}$ years the cavity shall move with respect to the AGN a few kpc or less; assuming a pressure dependence in distance from the AGN to be $p_{0}\propto d^{-1}$, the expected expansion shall be of the order of $2\times 10^{-3}$ of its current radius. Demanding that the pressure be non-negative in the cavity and repeating the calculations of section 3, we find that the magnetic field must decrease.  Assuming mass conservation we conclude that about $5\times 10^{-4}$ of the magnetic flux will be reconnected. Thus a timescale for reconnection to convert a significant fraction of the magnetic field into heat and thus radiation is $10^{9}$ years, longer by two orders of magnitude compared to the rising time. 

\begin{figure}
	\centering
		\includegraphics[width=0.45\textwidth]{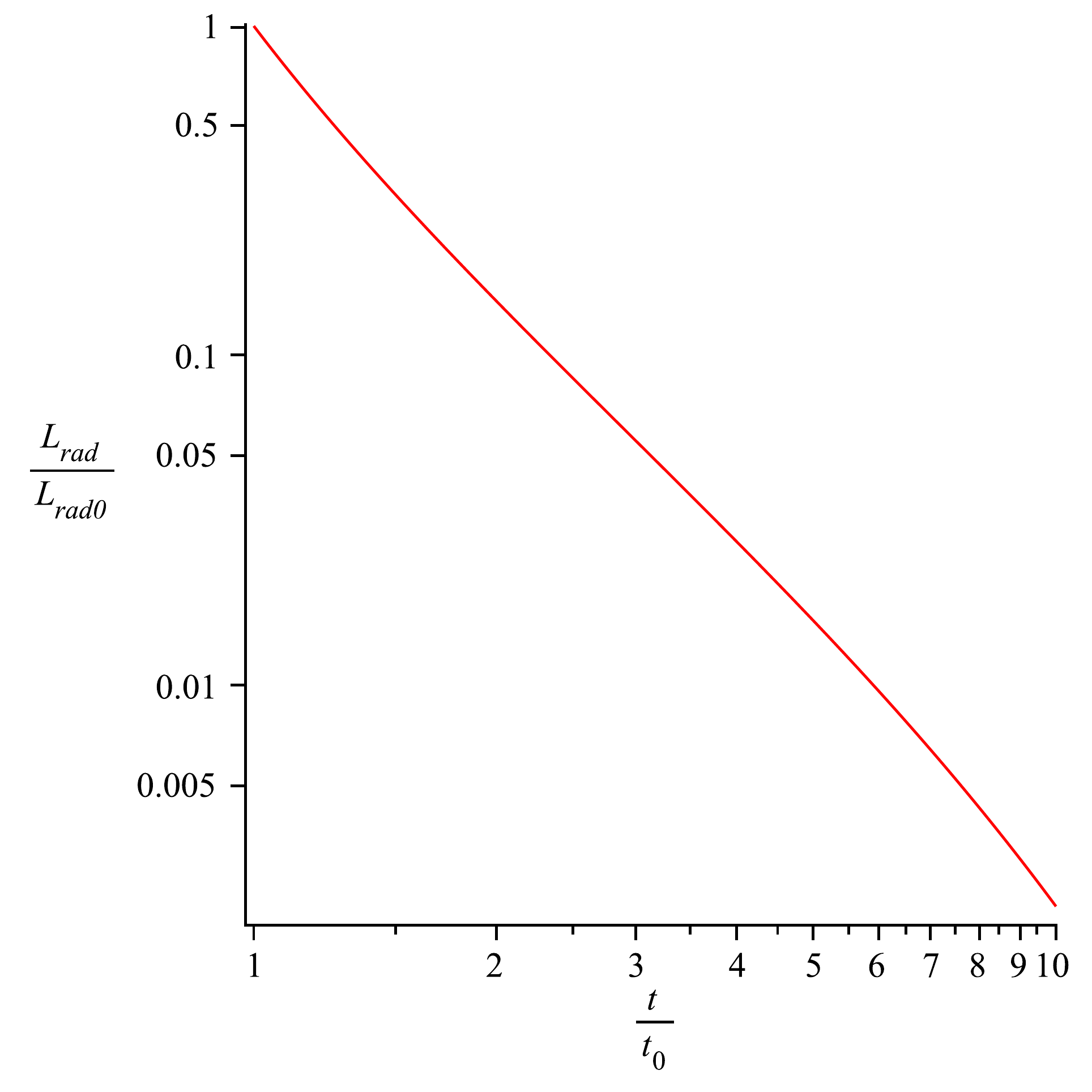}
		\caption{The radio luminosity expected by the cavity. While the cavity expands by a factor of a few, the luminosity drops almost three orders of magnitude, since the evolution of radio luminosity is $L_{rad} \propto t_{0}^{-3}$. }
		\label{RADIO}
\end{figure}

\begin{figure}
	\centering
		\includegraphics[width=0.45\textwidth]{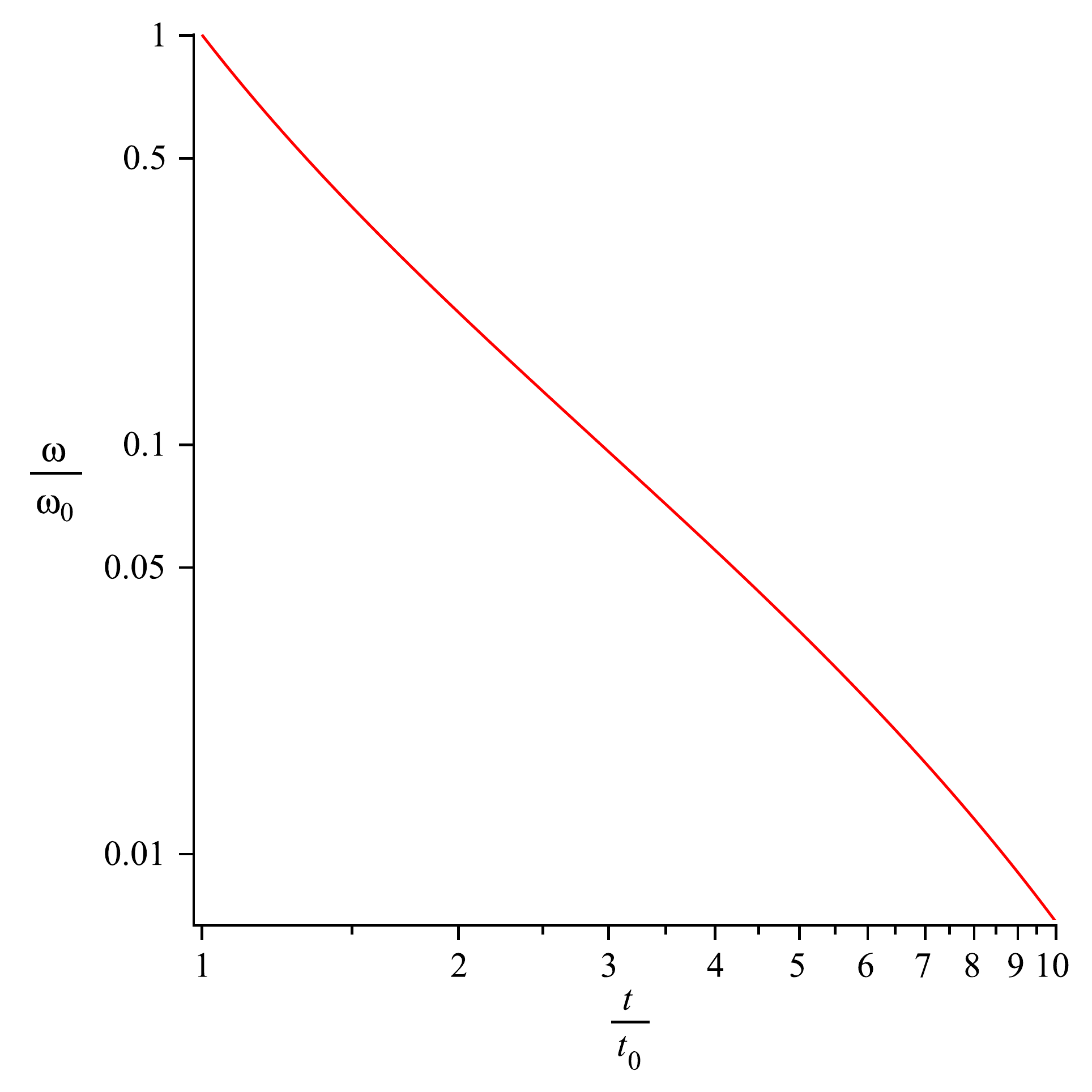}
		\caption{The peak frequency for synchrotron emission. We expect that it will move to much shorter frequencies as it moves away from the AGN, since its evolution is $\omega \propto t_{0}^{-2}$}
		\label{INDEX}
\end{figure}

A simple model describing the radio luminosity evolution is the following. Let us assume that the luminosity of the cavity is $L_{rad}=-Nm_{e}c^{2}\dot{\gamma}_{rad}$, where $N$ is the total number of emitting electrons in the cavity, and $\dot{\gamma}_{rad}$ is rate of change of the Lorentz factor, $\gamma$, because of radiative losses. Synchrotron emission depends on the magnetic field $\dot{\gamma}_{rad}\propto \gamma^{2}B^{2}$. $\gamma$ changes due to three factors: reconnection, which injects energy from the magnetic field in the plasma through heating, adiabatic loses to the intracluster medium through adiabatic expansion and radiative losses: $\dot{\gamma}=\dot{\gamma}_{rec}-\dot{\gamma}_{exp}-\dot{\gamma}_{rad}$. The timescale of expansion is $10^{7}$ years, the timescale for radiative losses is $10^{8}$ years and as we have shown above the reconnection timescale is even longer. We conclude that the radio luminosity evolution will be mainly affected by the adiabatic expansion, first through $\gamma$ and second through the magnetic field $B$. These two combined give the radio luminosity profile shown in Figure (\ref{RADIO}). The radio luminosity will evolve as $L_{rad}\propto t_{0}^{-3}$, where $t_{0}$ is the rising timescale.  The frequency where synchrotron radiation peaks is $\omega_{max}\propto \gamma^{2}B$, from the evolution of cavities we find that $\omega_{max} \propto t_{0}^{-2}$, Figure (\ref{INDEX}), making them harder to detect in radio at great distances from the AGN. This is consistent with the fact that the cavities that have a significant radio luminosity are the ones close to the AGN or directly connected to the jet, which is more evident when the evolution is studied in cavities of a given cluster \citep{Dunn:2005}. 

Thus, although explosive reconnection can produce large potentials that accelerate cosmic rays, it does not provide an efficient way of converting magnetic energy into heat and consequently radio emission. The main reason is that while reconnection per se is fast, the flux available to reconnect is limited and determined by the expansion rate which is much slower. From a different viewpoint this result can be understood in the following terms.  If the fraction of the flux comprising the reconnection layer over the total flux is $\delta$ and the volume of this area is $V_{rec}$ the energy available for radio emission is proportional to $V_{rec}^{-1/3}\delta^{2}$, whereas the potential that accelerates cosmic rays is proportional to $V_{rec}^{-2/3}R_{rec}\delta $. In addition even if the volume of the reconnection layer is small, one of its dimensions $R_{rec}$ can be large, i.e. of the order of the circumference of the cavity, thus the energy available for acceleration is $2\times 10^{20}{\rm eV}$ derived in the previous section and therefore sufficient to accelerate UHECRs.  

 \section{Discussion}
  
In this study we have found analytical solutions for expanding cavities, both for fully dynamical evolution and also in the quasi-static regime. The cavity is confined by some pressure that decreases with time, but remains constant around the cavity. In a more realistic environment and given that the cavity size is comparable with the distance the pressure of the IGM drops noticeably, we expect a deviation from spherical geometry. In addition, the drag force on the top surface may cause further deformation. The parametrization of such factors can be treated more appropriately through a numerical model and is beyond the scope of an analytical model such as ours. Another issue is the stability of the rising cavity. GBL have shown that static cavities are stable using both analytical and numerical methods. In case of strong drag forces on the rising cavity, we expect more significant perturbations. 

While the absence of surface currents makes less vulnerable to resistive decay, for appropriate values of $\Gamma$ current sheets will form and their equilibrium will not be described by that solution. This allows us to propose that rising magnetized cavities are UHECR accelerators. Expansion leads to lower density, and if the gas pressure of the cavity drops faster than the magnetic pressure the final outcome will be a region containing only magnetic field and confined by current sheets. It is those currents sheets that trigger resistive instabilities and allow the available  electric potential to accelerate cosmic rays. Order of magnitude estimations suggest that this mechanism is viable. 

\section*{Acknowledgements} 
This study is supported by NASA NNX09AH37G. The authors are grateful to Dimitrios Giannios for insightful comments and to Eric Clausen-Brown for comments on the manuscript.

\label{lastpage}

\bibliographystyle{mnras}
\bibliography{BibTex.bib}

\end{document}